\documentclass[useAMS,usenatbib]{mn2e_letter}
\usepackage[]{graphicx}
\usepackage{psfig}
\usepackage{epsfig}
\title[Characterization of hexabundles: Initial results]{Characterization of hexabundles: Initial results}

\author[J. J. Bryant et al.]{J. J. Bryant$^{1}$\thanks{E-mail: jbryant@physics.usyd.edu.au (JJB)}, J. W. O'Byrne$^{1}$, J. Bland-Hawthorn$^{1,2}$ and S. G. Leon-Saval$^{1,2}$\\
$^{1}$ School of Physics, The University of Sydney, NSW, Australia 2006; \\
$^{2}$ Institute of Photonics \& Optical Science, The University of Sydney, NSW, Australia 2006; \\
}

\begin{document} 
\maketitle 

\label{firstpage}

\begin{abstract}
New multi-core imaging fibre bundles -- hexabundles -- being developed at the University of 
Sydney will provide simultaneous integral field spectroscopy for hundreds of
celestial sources across a wide angular field. These are a natural progression
from the use of single fibres in existing galaxy surveys.  Hexabundles will allow us to 
address fundamental questions in astronomy without the biases
introduced by a fixed entrance aperture. We have begun to consider instrument concepts
that exploit hundreds of hexabundles over the widest possible field of view.
To this end, we have compared the performance of a 61-core fully-fused hexabundle 
and 5 lightly-fused bundles with 7 cores each. 
All fibres in the bundles have $100\mu$m cores.
In the fully-fused bundle, the cores are distorted from a circular shape in order to achieve a 
higher fill fraction. The lightly-fused bundles have circular 
cores and five different cladding thicknesses which affect the fill fraction.
We compare the optical performance of all 6 bundles and find that the 
advantage of smaller interstitial holes (higher fill fraction) is outweighed
by the increase in modal coupling, cross-talk and the poor optical performance caused by the deformation of 
the fibre cores. Uniformly high throughput and low cross-talk are essential for 
imaging faint astronomical targets with sufficient resolution to disentangle 
the dynamical structure. Devices already under development will have between 
100 and 200 lightly-fused cores,
although larger formats are feasible. The light-weight packaging of hexabundles is sufficiently
flexible to allow existing robotic positioners to make use of them.
\end{abstract}


\section{Introduction}
\label{intro}

Current and planned cosmological surveys in the optical and infrared 
have fundamental limitations. Multi-fibre [e.g. 2dF \citep{Col01} and SDSS \citep{Yor00}] and
multi-slit [e.g. DEIMOS \citep{Fab03} and VIMOS \citep{LeF03}] surveys 
have or will amass large catalogues of galaxies in
order to
deduce their global properties. However these suffer from biases introduced
when a fixed angular size aperture such as a single fibre is used 
to observe galaxies, irrespective of the size, distance or morphology \citep[see 
for example, fig.8 in ][]{Ell05}.
On the other hand, integral field units (IFUs) like TEIFU \citep{Mur00} and 
GMOS \citep{All97} or image slicers such as 
SINFONI \citep{Eis03} and NIFS \citep{McG03} spatially 
sample the spectra, giving morphological and dynamical information. However, 
current IFUs are restricted in the number of objects that can be observed. The ideal is to combine multi-object
spectroscopy (MOS) positioning technology with IFUs. Hexabundles can do
exactly that.

Hexabundles have up to many hundreds of closely-packed fibres to allow spatially
resolved spectroscopy. 
They have several practical advantages. Firstly, they are not as sensitive as single-fibre devices to seeing 
and positioning errors and are unaffected by seeing losses.
Secondly, they can be used with AO-corrected or natural seeing and
in the optical or infrared. Thirdly, there is no need for microlens arrays, and the 
plate scale can be changed simply with a single macro lens. 
Fourthly, they can be used with conventional fibre positioning
technology as imaging bundles to obtain spatial information in a survey of many thousands of galaxies.
The resulting scientific gains over existing large galaxy surveys would then include
the ability to investigate AGN triggering and feedback
including outflows, galaxy merger rates and merger-induced processes, the
substructure in
gravitational lenses, stellar populations and abundance gradients, 
as well as tracing the build up of dark matter, stellar mass
and angular momentum in galaxies. Decomposition of bulge and disk components 
would be possible for thousands of galaxies.
Hexabundles are particularly effective for galaxies with asymmetries, 
multiple components, mergers and
substructures including high redshift galaxies that are not yet formed
into spheroids or disks but consist of clumpy merging components which 
would be poorly sampled with a single fibre spectrum.
Differential binning of the fibres can also be applied, particularly for more distant
objects where the brightness decreases rapidly away from the nucleus.

A 100,000 galaxy survey out to a redshift of 0.2 will be possible with 
hexabundles in the 
proposed FIREBALL instrument on ESOs Very Large Telescope (VLT). Currently
the FLAMES instrument uses the OzPoz positioner to place 132 single fibres
across a $24'$ field, which then feed into the GIRAFFE spectrograph. The
proposed FIREBALL upgrade would involve fifty hexabundles, with around 
100-cores each 
sampling $\sim0.42''$.
 
Hexabundles are not limited in the number of cores in each bundle, beyond
the physical size of the bundle holder. So far we have made bundles up to 61
cores, but larger devices of several hundred cores would be suitable for
multi-object positioners. In anticipated applications, the core will have
$0.4 - 0.7''$ per core. Then 61-core and 367-core bundles would
have imaging areas $4-6''$ and $9-15''$ across respectively. 

The main trade-off with the design of hexabundles is
to have the largest fill-fraction (ratio of core area to total bundle
area) possible without compromising the
optical performance. The fill-fraction is affected by both the cladding thickness
and how the fibres are packed together. Fibres that are fully-fused together
have smaller gaps between cores giving a fill-fraction of over 90\%, but are 
significantly distorted from a circular shape. On the other hand, lightly-fused 
fibres remain circular but have larger gaps between them, resulting in 
fill-fractions less than 87\%.

This paper describes the performance tests of our first fully-fused and lightly-fused 
hexabundles with an aim to assess the trade-offs of fill-fraction versus 
optical performance.  
Section 2 describes the hexabundle devices, while
the experimental method and data reduction are in section 3. Sections 4 and 5 
have
the results for the fully-fused 61-core hexabundles and the lightly-fused 
7-core hexabundles respectively. Final conclusions on the comparison of the
different types of hexabundles is in section 7.

\section{Hexabundle devices}

\begin{figure}
\centerline{\psfig{file=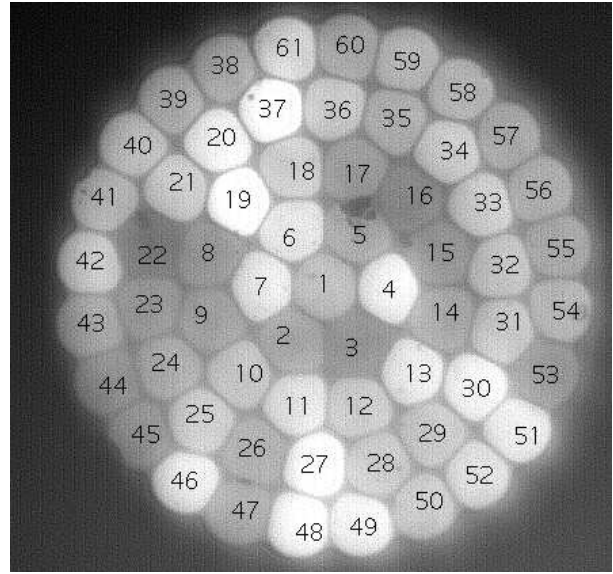, width=8.0cm}}
\vspace*{3mm}
\caption{An image of the 61-core fully-fused hexabundle taken with non-uniform 
illumination to show the shape and position of the cores. Cores 
$1-16$, $17-31$, $32-46$ and $47-61$ were grouped together at output 
on four glass plates. 
Each
of the cores are $100\mu$m in diameter. The cores are distorted from 
circular in order to increase the fill-fraction.
}
\label{61bundle_pic}
\end{figure}

Six new hexabundle devices have been characterized. In each device, one end 
has multimode fibres with reduced cladding thickness, fused into a single element, which is the hexabundle. The fibres at the other end are loose and have normal cladding thickness. 
The first of these hexabundles consists of 61 fully-fused fibres without interstitial holes (Fig.~\ref{61bundle_pic}) giving
a fill-fraction of $>90$\%. 
However, all but the central fibre are significantly distorted from a circular shape. The
fibres are 4.5\,m long and each fibre had core and cladding diameters of $105\mu$m and 
$125\mu$m respectively, before the cladding was etched away over a $2-3$\,cm length where the 
fibres were then fused \citep[further details of the manufacturing 
process can be found in][]{Bla10}. The resulting hexabundle has a diameter 
of $\sim900\mu$m. 
On 
output, the 61 fibres are lined up in V-grooves between glass plates with 
three rows of 15 and one row of 16 fibres. The fibres in adjacent grooves in the same glass 
plate are separated by $500\mu$m.

\begin{figure}
\centerline{\psfig{file=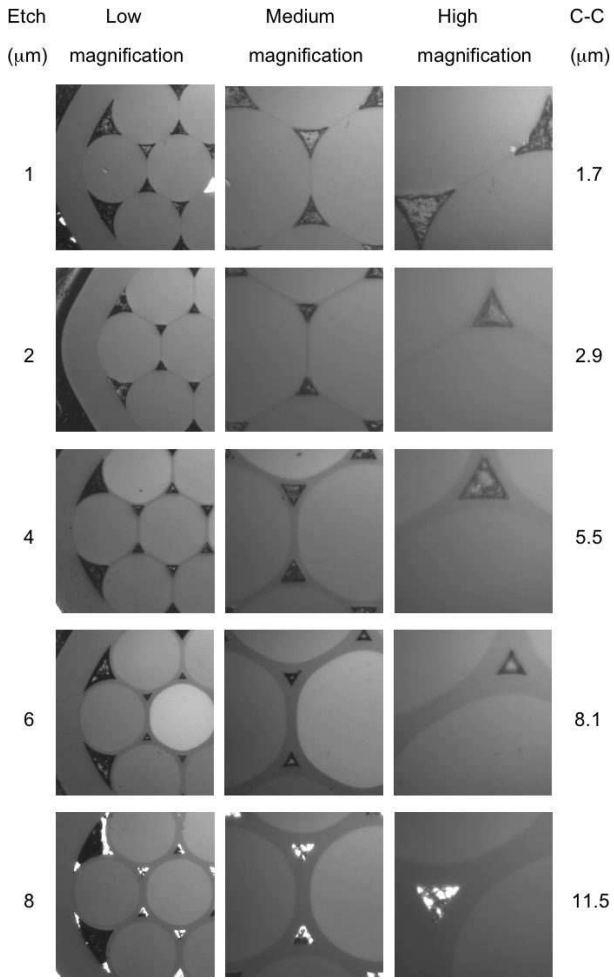, width=8.0cm}}
\vspace*{3mm}
\caption{
The 7-core lightly-fused hexabundles, shown with different magnifications. 
In each horizontal panel, the cladding thickness after etching away part the cladding is
listed to the left
and the core-to-core separation is listed to the right. 
The images show the bundles before the interstitial holes were filled with 
soft, low refractive index glue.
In each
of these bundles the cores are $100\mu$m diameter. These lightly-fused 
bundles are each the same size as the central 7      
cores of the 61-core bundle shown in Fig. ~\ref{61bundle_pic}.} 
\label{lightly-fusedbundle_pic}
\end{figure}

The remaining five hexabundles have 
seven lightly-fused fibres (with small interstitial holes) with 
cladding thicknesses of 1, 2, 4, 6 and $8\mu$m. The main difference to the
previous bundle is that the fibres are circular rather than distorted in shape 
(see Fig.~\ref{lightly-fusedbundle_pic}),
 resulting in lower fill-fractions. 
In these bundles, the output
fibres are loose and were mounted into 35 parallel V-grooves on a plate
with careful attention not to apply pressure to the fibres.
We aim to investigate the effect of circular cores on the optical performance and
consider the trade-off with fill-fraction. 
The different cladding thicknesses are used to assess the ideal balance between
minimising cross-talk while maximising
fill-fraction.

\section{Experiment method and data reduction}

\begin{figure}
\centerline{\psfig{file=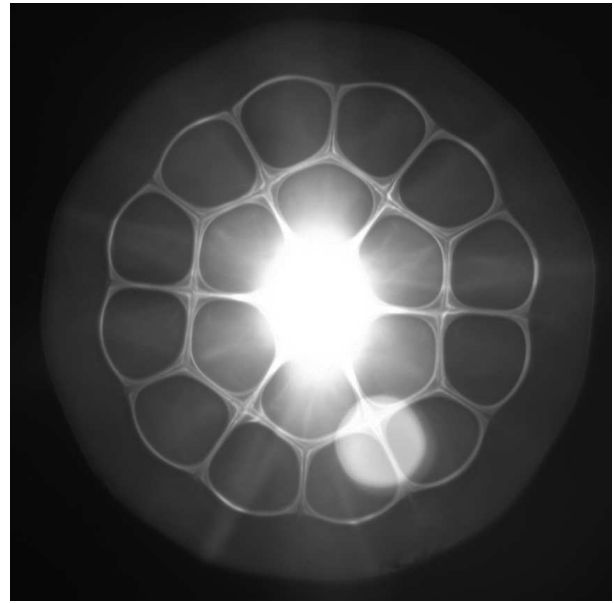, width=8.0cm}}
\vspace*{3mm}
\caption{This picture of an earlier prototype bundle (Skovgaard, private 
communication) shows that when coherent light is input into the output multi-mode fibres 
of a hexabundle, the light coming out of the hexabundle has a speckle 
pattern that at best focus, appears larger than the core of the fibre. The spot
at the lower right is an obvious reflection. }
\label{speckle}
\end{figure}

Each of the bundles were tested for cross-talk, relative throughput and mode-dependent losses.
It was crucial for the cross-talk tests in particular,
to ensure that no stray light was getting into any core other
than the intended target.
Originally laser light was focussed into the individual fibre (output) 
end of the fibres while the hexabundle end was imaged with the camera. 
This has the advantage that the fibres could be separated 
so that no stray light from the laser could go into adjacent 
fibres. However, the light coming out of the multimode 
fibres at the hexabundle has a speckle pattern that is larger than the size 
of the core. This is because the different modes focus in slightly different 
planes outside of the polished bundle face so that all of the light out of the 
fibre cannot be focussed at the same time, resulting in an imaged 
spot size that has an apparent diameter of approximately 125\% of the core 
diameter. An example of this effect is shown in Fig.~\ref{speckle}. In any
adjacent fibre it is impossible to distinguish light from this oversize speckle pattern from light leaked into the fibre through cross-talk.
 
\begin{figure*}
\psfig{file=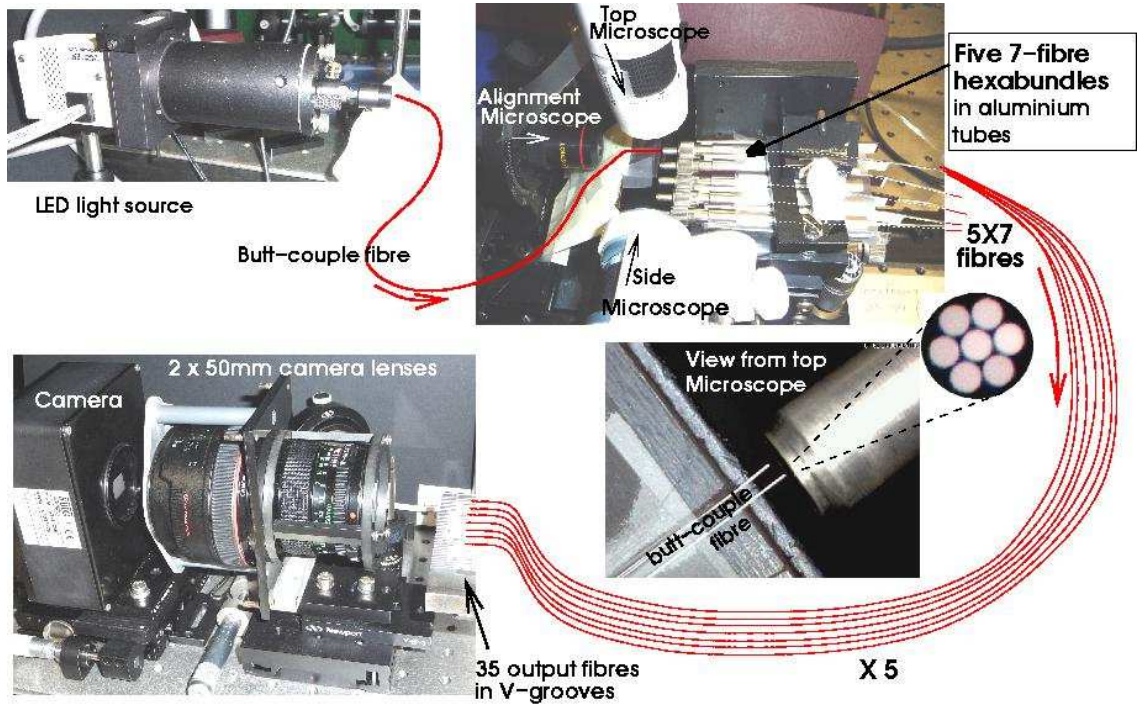, width=15.0cm}
\vspace*{3mm}
\caption{Equipment setup to test the hexabundles. Clockwise from top left: 
The light from an LED source passes through a filter holder and is focussed 
into a $50\mu$m core fibre. This fibre is butt-coupled in turn to each
core of each hexabundle. The five metal tubes lined up are the lightly-fused 7-core 
hexabundles (the fully-fused 61-core hexabundle was in the same position when
being tested). An
xyz stage allows the bundle cores to be accurately aligned with the 
butt-couple fibre. An 
alignment microscope allows us to see which hexabundle core the input fibre 
is butt-coupled to.
Top and side microscopes are used to check the butt-couple fibre is
within $100\mu$m of the bundle face. The view from the top microscope
is inset, showing the butt-couple fibre aligned with a hexabundle. 
The second butt-couple fibre in this image was a spare, and not used. 
Each of the 5 bundles has 7 fibres
coming out, giving 35 fibres mounted in V-grooves 2\,mm apart on an xyz stage. However,
when the fully-fused bundle was being tested, the 61 output fibres were aligned
on four glass plates on the xyz stage. The
output fibres are imaged through a pair of matching lenses onto a camera.}
\label{setup_pic}
\end{figure*}

The final method adopted, as shown in Fig.~\ref{setup_pic}, used an Oriel LED light source which was 
focussed into a $50\mu$m core fibre with a $60\mu$m diameter 
including cladding. This fibre was then butt-coupled to the 
$100\mu$m hexabundle cores (reversing the direction of the light through the
fibres compared to the previous method - this is the direction the hexabundles will be used in 
astronomical applications). 
We mounted the butt-couple (input) 
fibre in a V-groove on a narrow plate that was attached to a tilt/rotation 
stage off to the side of the hexabundle mount. This minimised obstruction of 
the area around the hexabundle face so that the fibre could be adjusted to align perpendicular to the
hexabundle face. 
The narrow plate could then slot in from the side between the hexabundle and 
a microscope with the fibre
bending abruptly away to the side. This allowed the microscope to be positioned close enough to 
image the hexabundle and input 
fibre at high enough magnification to be sure the input fibre was centred on
a chosen fibre core within the hexabundle.
The input fibre needed to be less than $100\mu$m 
from the hexabundle face to prevent light straying into adjacent
fibres. 
The small outer diameter of the butt-coupled fibre was essential in 
order to not obscure the view of the hexabundle cores.
Two other
microscopes mounted above and side-on were used to ensure that the fibre was
perpendicular to the hexabundle face in both axes.
The maximum error contribution to the NA from the input alignment
was $\pm0.002$.
The output fibres were each imaged through a pair of large camera lenses onto
an SBIG camera. A micrometer stage allowed the distance from the focus to
the backfocus position of the 
camera to be accurately measured.

For all the measurements, the LED source light level was kept constant. Each
measurement was done twice, once with a Bessel blue filter (centred on $0.45\mu$m) then with 
a Bessel red filter (centred on $0.65\mu$m)
between the LED source and the focussing assembly.
Time variation in the input light intensity plus variations in the
SBIG camera response, were quantified with repeated images through one
core of the bundle at time periods ranging from second to hours. The 
maximum variation in resulting integrated counts was $<1.0$\% with typical
values of $\sim0.2$\%.

The relative throughput of every fibre in each bundle was measured by centring the 
input light on each fibre.  For cross-talk measurements, 
the light was centred on a core while 
the surrounding fibres
were imaged.  FITS images from the SBIG camera were processed using the {\sc iraf} 
astronomical data reduction software \citep{Tod86}. Firstly each image was flat fielded.
The background was
fitted and subtracted in {\sc iraf radprof} using an annulus with inner radius
approximately 3 times the maximum radius of the fibre image.
Then the total integrated counts were measured in {\sc iraf radprof} by 
fitting the barycentre of the fibre image and selecting a fixed aperture size
for all fibres that was large enough to include all the output from the fibre.
The target fibre and the surrounding fibres were measured through the same
size aperture.

\begin{figure*}
\psfig{file=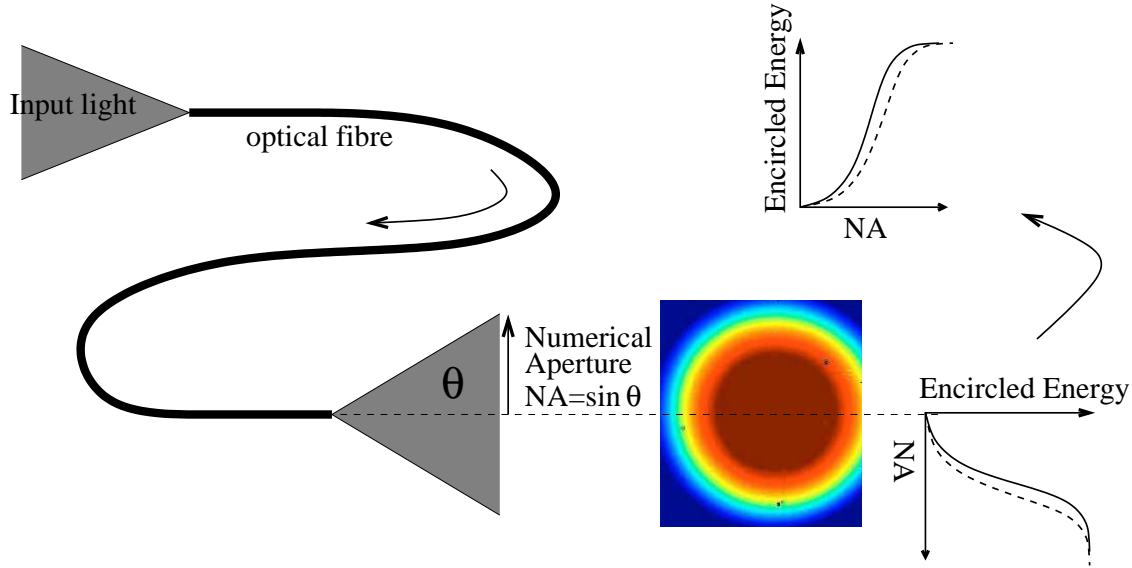, width=15.0cm}
\caption{FRD increases the output cone angle $\theta$. Therefore
the comparative FRD in the fibres within a bundle can be assessed from
plots of encircled energy vs numerical aperture (NA), where NA is sin$\theta$.
Worse FRD shifts the profile to the right (dashed line). 
When more
light is coupled to higher-order modes, more of the encircled energy is at
higher NAs, such that the
shape of the encircled energy profile will broaden or shift to the right.
}
\label{NAEEdiag}
\end{figure*}

We also aimed to compare the performance of different fibres within the same
bundle in terms of focal ratio degradation (FRD) or mode-dependent losses.
FRD (e.g. Carrasco and Parry 1994) is a loss of optical entropy or non-conservation of {\it \'{e}ntendue} due
to mode mixing as light propagates down the fibre.
This modal mixing causes the opening angle of the output 
beam to be larger than that of the input beam resulting in a wider encircled energy distribution (see Fig.~\ref{NAEEdiag}).
The modal coupling means that lower-order modes will ``couple" to higher-order 
modes which have a larger angle at the output beam, hence a larger NA.
In order to compare the consistency of the performance
of different fibres within the same bundle, it was therefore important
to ensure that all modes were filled on input. 
When the cores are distorted,
under-filling the NA on input may mean the light couples into different
modes at the input for different fibres, making it invalid to compare the
distribution on output. Therefore we have filled the NA
of the fibres to excite all possible modes, so that the output energy distribution across the output 
image in the back-focal plane can be directly compared for each fibre. 
The encircled energy profile 
at back-focus, will therefore be broader for higher modal coupling.

The camera was shifted to a known back-focal
distance; selected to give sufficient pixels across the output fibre image 
to accurately fit an encircled energy profile. Using the known pixel size of the camera, an NA value was 
calculated from sin\,$\theta$ of the cone angle. This gives the 
effective output NA at each point across the image of the fibre. This NA value
is used in the graphs to follow.
After flat fielding, {\sc iraf radprof} was used to fit the centre 
position along with the background and 
peak counts in each image profile. These initial values were then input into 
{\sc iraf pprofile}, and the radially-collapsed encircled energy profile was 
fitted interactively. 
An iterative procedure was used to determine the background value to subtract 
and the peak counts in the profile until the background was flat and the encircled 
energy levelled at 100\%. 
The encircled energy values could then be matched to the NA calculated for each
radial pixel position in the image.
The resulting plots of NA vs encircled energy will therefore be shifted
to the right when there are worse mode-dependent losses 
(larger mode-dependent losses will give higher FRD).

\section{61-core hexabundle performance}

\subsection {Encircled energy profiles}
\begin{figure*}
\centerline{\psfig{file=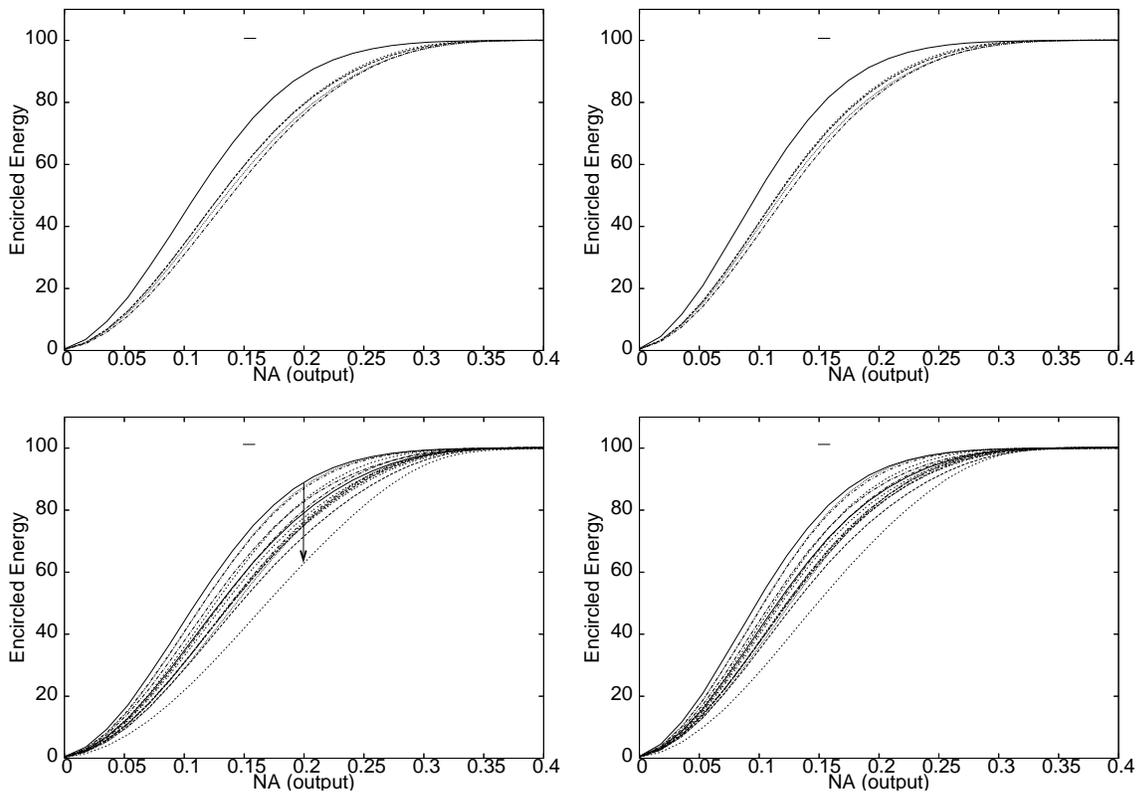, width=15cm}}
\vspace*{3mm}
\caption{{\it Fused fibre bundle results.} (Top) The encircled energy vs numerical aperture (NA) at output for
the central fibre (solid line) and
the average of all the core numbers $1-16$, $17-31$, $32-46$ and $47-61$ 
(short dashed, long dashed, dotted, dash-dot respectively). 
(Bottom) The same profiles but for the 
individual fibres $1-16$, to show the scatter around
the average values.
The plots on the left are through the blue ($0.45 \mu$m) filter and those on
the right are through the red ($0.65 \mu$m) filter. The arrow indicates  
how we defined encircled energy drop from the central fibre to the worst fibre.
A short horizontal line at NA=0.15 indicates the
measurement error in the profiles at that NA, primarily
due to focussing the output from each multimode core onto the camera.
}
\label{avEE}
\end{figure*}

The encircled energy profile of each of the fibres in the 61-core hexabundle, was measured against
the output NA. 
Fig.\ref{avEE} shows the encircled energy profiles through both the 
Bessell blue ($0.45 \mu$m) and red ($0.65 \mu$m) filters. The profiles 
through the red filter
are at a lower NA than those through the blue filter because the shorter wavelengths have
more modes and therefore a higher NA.  While the
average profiles of 16 cores get worse as the range of core numbers gets further from
the bundle centre, 
there is a very large variation in encircled energy profiles from 
fibre-to-fibre, which will give inconsistent performance across the bundle
for imaging and spectroscopy.

In order to quantitatively compare the encircled energy profiles,
the drop in encircled energy compared to the central fibre was calculated
for a nominal NA of 0.20. This NA is chosen because it is less than the total size of the output
cone (output NA at 100\% encircled energy). The maximum NA is truncated by 
mode-dependent losses out of the fibre, however the 
distribution of encircled energy within that cone will also be broadened by 
modal coupling, and more modal coupling is indicative of 
more FRD. Therefore, by selecting the encircled energy loss at an NA of 0.2,
we are measuring the broadening of the profile due to modal coupling.
The resultant drop in encircled energy ranges
from less than 1\% up to more than 25\% for different fibres in the bundle.
At 90\% encircled energy, the up-conversion of NA is 0.10 for the worst fibre 
in both the blue and red filters.

To see if the deformation of each fibre affects the 
encircled energy profiles, the deformation was measured by fitting circles
to the image shown in Fig.~\ref{61bundle_pic}. The largest circle that
does not include any cladding, and the smallest circle that includes all the
core were fit to each core. The ratio of the diameters of these circles
was used to define a `deformation ratio'. While this measure is subjective,
in Fig.~\ref{EEdrop}, the plot for deformation ratio against encircled 
energy drop shows a correlation with Spearman 
rank correlation coefficient of 0.34 (95\% probability).
Therefore, the fibres with the highest deformation ratio tend to have the largest 
encircled energy drop, indicating the highest modal coupling.

\subsection{Cross-talk and throughput}

Cross-talk was measured for 14 random fibres in the bundle by measuring the
integrated counts in each adjacent fibre while the light was centred on the 
target fibre. The loss was not equally distributed into the adjacent fibres, 
with most target fibres showing the majority of the cross-talk into
one or two of the adjacent fibres while in some cases the counts were more
than 100 times lower in other
adjacent fibres. This was due partly to significant differences in the contact
area and cladding width between different fibres. Such differences arise from
the distortion of the fibres in the fully-fused bundle. 
The 
total loss into adjacent fibres was predominantly in the range of $0.11-1.2$\% (0.005-0.06dB) 
with one fibre at 4.2\% with the blue filter, and one at 4.4\% (0.2dB loss) in 
the red filter. The total cross-talk could not be evaluated in the outer 
row of the hexabundle because we have shown that the loss is unevenly distributed to adjacent
fibres and therefore could not be assessed where there were no adjacent fibres 
on the outer edge. As the outer row of fibres include some of the most distorted profiles, the
total cross-talk is likely to be above 0.2dB for some of those fibres. For 
example, the cross-talk from fibre 56 into just 55 and 57 is already a total of 0.17dB 
alone and therefore would be much higher if the losses could be 
measured on all sides.

\begin{figure*}
\centerline{\psfig{file=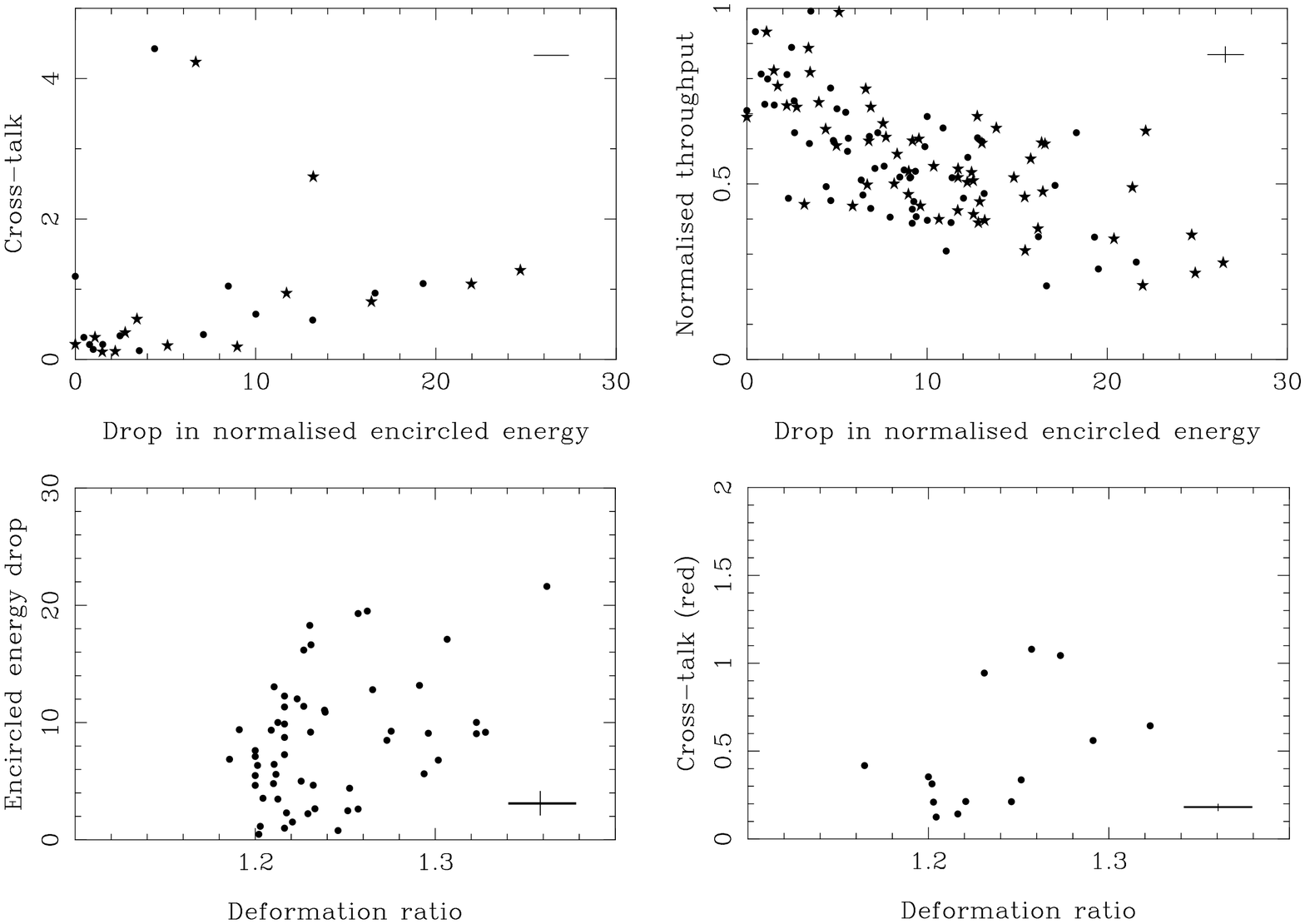, width=15cm}}
\caption{{\it Fused fibre bundle results.} Drop in percentage encircled energy compared to the 
central fibre vs total cross-talk (top left), throughput (top right)
and deformation ratio (bottom left), along with deformation ratio vs cross-talk (bottom 
right). In all panels the dot and star symbols are for the $R-$band ($0.65\mu$m) and
$B-$band ($0.45\mu$m) respectively. The maximum errors  
are shown by the error bars on each plot. 
In the top left plot, the cross-talk errors are 
smaller than the point sizes in all but the 2 highest points, where it is 
slightly larger than the point. 
}
\label{EEdrop}
\end{figure*}

In Fig.~\ref{EEdrop} there is an anti-correlation (with Spearman-rank test 
correlation coefficient of -0.65, and 99.9\% probability) between the drop in encircled 
energy and the throughput.
Throughput losses are a combination of higher cross-talk in the more distorted
fibres and modal stripping (indicative of worse FRD). The encircled energy drop is
not measured at the edge of the output light cone but at NA$=0.2$. FRD causes
lower-order modes to be coupled to higher orders, broadening the beam
to give this encircled energy loss. At the same time, the higher-order modes
can be lost entirely and therefore the throughput is affected by worse modal 
stripping.
The performance of different fibres
within the same bundle is very inconsistent, making it ineffective as an imaging device. 
With the exception of one fibre with very high cross-talk, that
performed differently to the others in every test, the encircled energy drop was
also correlated with cross-talk with a Spearman Rank correlation
coefficient of 0.60 (with $>99$\% probability). 
It can also be seen that the throughput varied by a factor of five between each of the 61 cores which would have an impact on the accuracy of photometry 
in extended astronomical sources. 
It is clear that the more distorted fibres have lower throughput,
worse cross-talk and higher encircled energy drop.

\subsection{Cladding scatter}
\label{cladscat}

It was also noticed that when the input light was directed at the cladding
between fibre cores, that light was then dispersed through the entire bundle.
In an astronomical application, the component of an extended source
that hit the cladding would result in all positional information being lost
and the spectrum of that component of the object being added to every fibre
across the field. By centring the light on the cladding between fibres 11 and
12, then measuring the counts in 6 fibres that were not adjacent, 
the counts were $0.1-0.2$\% of the counts in fibre 12 when the light was 
directed into fibre 12. Fibre 12 was chosen because the throughput counts in that fibre 
are average compared to the 
counts in each of the other 61 fibres. Based on an estimate of the cladding
area covered by the light in this test compared to the total cladding
area, up to 25\% of the light in any one fibre could be scattered from 
cladding elsewhere in the bundle. This fraction would lead to unacceptable 
confusion in the spectroscopic imaging of an extended astronomical source.

The original aim of fully-fusing
the bundle was to increase the fill-fraction to $>90$\% . However, based on this
performance, the gain in coverage across an object being imaged, is outweighed
by the confusion in the spatial origin of the light at the spectrograph due
to cross-talk and cladding scatter. Also, inaccuracies in photometry will
result from the extreme differences in throughput across an image.

The next step was to consider the improvement in this performance if the 
hexabundle was lightly-fused so that the fibres remain circular at the 
expense of a slightly smaller fill-fraction. 

\section{Performance of the 7-fibre non-fully-fused hexabundles}

We had five hexabundles made with cores that are lightly-fused. These have the advantage that the fibres are not distorted and
therefore is expected to perform more consistently.
In order to compare cross-talk for a given fill-fraction, they were made
with cladding thicknesses of 1, 2, 4, 6 and $8\mu$m \citep[see][for a 
discussion on the issues surrounding the manufacture of 
bundles with these cladding thicknesses]{Bla10}. The aim was
to test how thin the cladding could be before the cross-talk became too large.

\subsection{Encircled Energy profiles}

\begin{figure*}
\centerline{\psfig{file=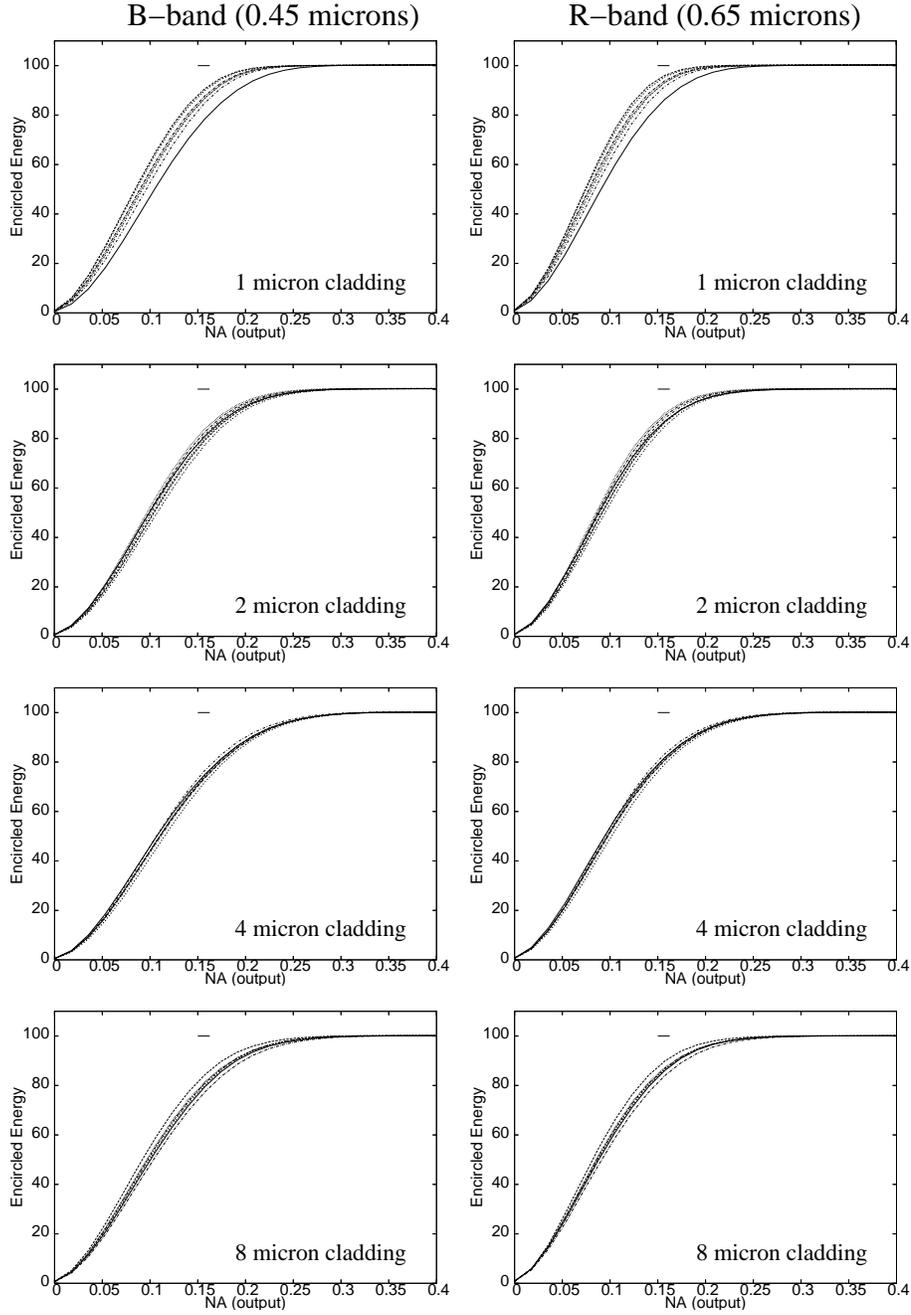, width=12.0cm}}

\vspace*{3mm}
\caption{ {\it Unfused fibre bundle results.}
The encircled energy vs numerical aperture (NA) at output for
the lightly-fused bundles with 1, 2, 4 and $8\mu$m cladding thicknesses. 
Results through the $B-$band and $R-$band filter are in the left and right
columns respectively. The solid line is the central core while the surrounding
cores are the dotted and dashed lines. A short horizontal line at NA=0.15 indicates the
measurement error in the profiles at that NA, primarily
due to focussing the output from each multimode core onto the camera.
}
\label{EElightly-fused}
\end{figure*}

Fig.~\ref{EElightly-fused} shows the encircled energy profiles for the lightly-fused 
bundles. The profiles of each of the 7 cores in the 2, 4 and $8\mu$m
clad bundles 
agree within errors, indicating consistent modal coupling performance across the bundle. 
This is in contrast to the large variation between cores in 
the fully-fused bundle (see Fig.~\ref{avEE}). 
Each of the profiles in Fig.~\ref{EElightly-fused} have lower NA than (sit to 
the left of) the
best performing fibre (central fibre) of the fully-fused bundle, indicating 
lower modal coupling (and therefore lower FRD) 
in the lightly-fused bundles.
The
$1\mu$m cladding bundle has a distribution of encircled energy profiles for the
7 cores that is slightly larger than errors, which may be due to
mode stripping induced by the higher cross-talk. In this case the central core has worse NA than
the other cores as would be expected from mode stripping based on the symmetry of the bundle.

\subsection{Cross-talk and throughput}

For each of the lightly-fused hexabundles, the relative throughput was measured for each
of the 7 fibres, and the cross-talk was measured by summing the losses into the 
surrounding 6 fibres when light was directed down the central core.

Table~\ref{FFvsCT} shows the cross-talk performance through the $B-$ and $R-$band
filters compared to the fill-fraction. While the hexabundles with 4, 6 and $8\mu$m
clad fibres had no cross-talk down to the limit we could measure in our
images, they also have a lower fill-fraction. The cross-talk was worse for the
$1\mu$m clad bundle than for the $2\mu$m clad bundle, as expected. 
In the anticipated uses of these hexabundles (e.g. FIREBALL) the fibres would
be $0.4-0.7''$/core - comparable to the best seeing at an excellent
telescope site of $0.4''$, or average seeing at a good
telescope site of $0.7''$. If the core is the size of the FWHM of a gaussian
seeing profile, then 50\% of the encircled energy will be in the central
core and $\sim45$\% in the adjacent the surrounding cores. The
$2\mu$m clad bundle has a total cross-talk of $0.39$\% (at $0.65\mu$m), which 
is $<1$\% of the 
star light in
the surrounding cores, but for the $1\mu$m clad bundle (with up to $4.8$\% cross-talk), approximately a tenth of the 
light in the surrounding cores will be cross-talk contamination. Therefore,
at optical wavelengths, the $2\mu$m clad bundle is suitable for applications where
the core size is comparable to the FWHM, and the $1\mu$m clad bundle may also
be suitable if there are several cores within the FWHM of the seeing profile. 
In the infrared, longer wavelengths will cause the cross-talk to be worse and
thicker cladding would be required.

When the light was directed at the cladding, there was no detectable
transmission of that light into all the cores as was seen with
the ``cladding scatter" in the fully-fused bundle (see section~\ref{cladscat}).

\begin{table}
\caption{Total cross-talk from the central fibre into all of the 6 
surrounding fibres is listed as both a percentage of the integrated 
counts in the central fibre and as a dB loss. Values are shown for the
blue $B-$band and the red $R-$band Bessel filters. The first column lists the 
cladding thickness, then the 
fill-fraction (ratio of core area to total bundle 
area) is given for each cladding thickness. Limits listed are 
set by the image detection thresholds in cases where no emission was detected.}
\label{FFvsCT}
\begin{center}       
\begin{tabular}{cccccc}
\hline
Clad & Fill  & \multicolumn{4}{c}{Cross-talk}\\
 & fract. & \multicolumn{2}{c}{$B$ ($0.45\mu$m)}  & \multicolumn{2}{c}{$R$ ($0.65\mu$m)} \\
($\mu$m) &        & \multicolumn{4}{c}{\% (dB in brackets)} \\
\hline
\hline
1 & 0.87 & & 1.4 (0.06) & & 4.76 (0.21) \\
2 & 0.84 &  &0.25 (0.011) & & 0.39 (0.017) \\
4 & 0.78 & & $<0.01$ ($<0.003$) & & $<0.01$ ($<0.003$) \\
6 & 0.72 & & $<0.01$ ($<0.003$) & & $<0.01$ ($<0.003$) \\
8 & 0.67 & & $<0.01$ ($<0.003$) & & $<0.01$ ($<0.003$) \\
\hline
\end{tabular}
\end{center}
\end{table}

For the two bundles for which the cross-talk was measurable above the noise limit, the 
variation in the cross-talk values for the 6 surrounding fibres
was at worst a factor of 4. This is significantly lower than the factor
of $>100$ we found for the
61-core fully-fused hexabundle. We attribute this to the uniformly
circular fibres in the lightly-fused bundles which have equal contact 
areas with adjacent cores.

\begin{figure}
\centerline{\psfig{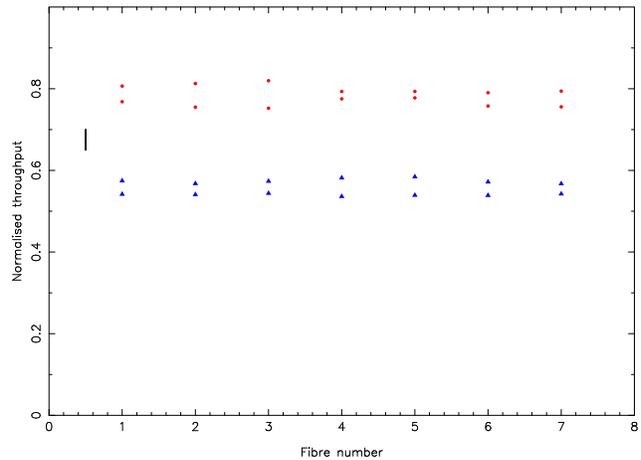}}
\vspace*{3mm}
\caption{ {\it Unfused fibre bundle results.} Normalised integrated counts (throughput) for each of the 7 
fibres in the hexabundles with $2\mu$m 
and $8\mu$m cladding thicknesses.
Results for the $R-$band and $B-$band are shown as dots and triangles respectively, and are normalised to different values to offset them for clarity. An error 
bar is given by the vertical black line. The variation in throughput between 
different fibres in the same bundle was less than the errors for both bundles shown.
}
\label{ThruNUM}
\end{figure}

The even, circular-nature of the fibres in the lightly-fused bundles has 
also resulted in a much more uniform throughput from fibre-to-fibre. In
Fig.~\ref{EEdrop} the variation in throughput between different fibres in 
the fully-fused bundle was up to $\sim500$\%. However, the lightly-fused 
bundles showed at worst 24\% variation between fibres in any bundle for
a given filter. Typically the variation was a lot less than this, as shown 
in Fig.~\ref{ThruNUM}. 
 
%
%

\section{Conclusion}

We have compared the performance of a 61-core fully-fused hexabundle to that of 5
lightly-fused bundles with varying cladding thickness. The distortion of the cores
in the fully-fused bundle was found to significantly increase cross-talk, reduce 
throughput and increase mode-dependent losses. The throughput varied by a factor of 5 between
different cores in the bundle, while the cross-talk reached over 4\%
in the worst core. The up-conversion in NA from the best (central) core to
the worst was 0.1 at 90\% encircled energy. The performance of the lightly-fused
cores was better in every respect, and we attribute that to the uniformly circular
cores. The throughput varied by at most 24\% and the modal coupling results were consistent
between the 7 cores in each bundle within errors. After comparing the 
cross-talk to the cladding thickness, we found that $2\mu$m cladding
bundle had $<0.4$\% cross-talk in the $R-$ and $B-$bands while
still having a fill-fraction of 84\%. While the fill-fraction of the fully-fused
bundle is $>90$\%, the superior optical properties of the lightly-fused 
bundles outweighed the gain in fill-fraction.

\section*{Acknowledgements}

We would like to thank Roger Haynes for valuable discussions and input.
JJB and JBH are 
supported by ARC grant FF00776384 which supports the Astrophotonics programme at the
University of Sydney.

\end{document}